# Is the Cosmic "Axis of Evil" due to a Large-Scale Magnetic Field?

Michael J. Longo
University of Michigan, Ann Arbor, MI 48109

I propose a mechanism that would explain the near alignment of the low order multipoles of the cosmic microwave background (CMB). This mechanism supposes a large-scale cosmic magnetic field that tends to align the cyclotron orbit axes of electrons in hot plasmas along the same direction. Inverse Compton scattering of the CMB photons then imprints this pattern on the CMB, thus causing the low-$\ell$ multipoles to be generally aligned. The spins of the majority of spiral galaxies and that of our own Galaxy appear to be aligned along the cosmic magnetic field.

Results from surveys of the cosmic microwave background, notably the Wilkinson Microwave Cosmic Anisotropy Probe (WMAP) [1], show remarkable agreement with the simplest inflation model. However, this tidy picture is marred by anomalies in the low order multipoles of the distribution. These include a near alignment of the quadrupole and octopole axes and a suppression of power in the low-$\ell$ multipoles. [See, for example, Refs. 2,3.] The probability that these alignments could happen by chance is estimated to be <<0.1%. [3,4] This bizarre alignment has been dubbed "the Axis of Evil" (AE). [4] The dipole is also found to lie along the same direction. Since the dipole is commonly assumed to be due to our motion through the rest frame of the CMB, the alignment of the dipole was thought to be a coincidence.

Campanelli *et al.* [5] propose that the power suppression in the low-$\ell$ multipoles can be explained if the universe is "ellipsoidal" with an axis generally along the direction of the quadrupole. They suggest that the eccentricity is produced by a cosmic magnetic field $B_0 \simeq (4-5) \times 10^{-9}$ G.

I show here that a uniform [6] cosmic magnetic field alone can explain most of these anomalies without the need for an ellipsoidal universe. The effect of a large-scale magnetic field is to cause the thermal velocities of electrons to move collectively in cyclotron orbits around the field lines, leading to a large-scale correlation in their orbital axes. The inverse Compton scattering of the electrons on CMB photons then produces an imprint of the magnetic field in the microwave power as illustrated in Fig. 1. The multipole moments of the magnetic field are thus mapped into the CMB, all of which are generally in the same direction as the dipole.

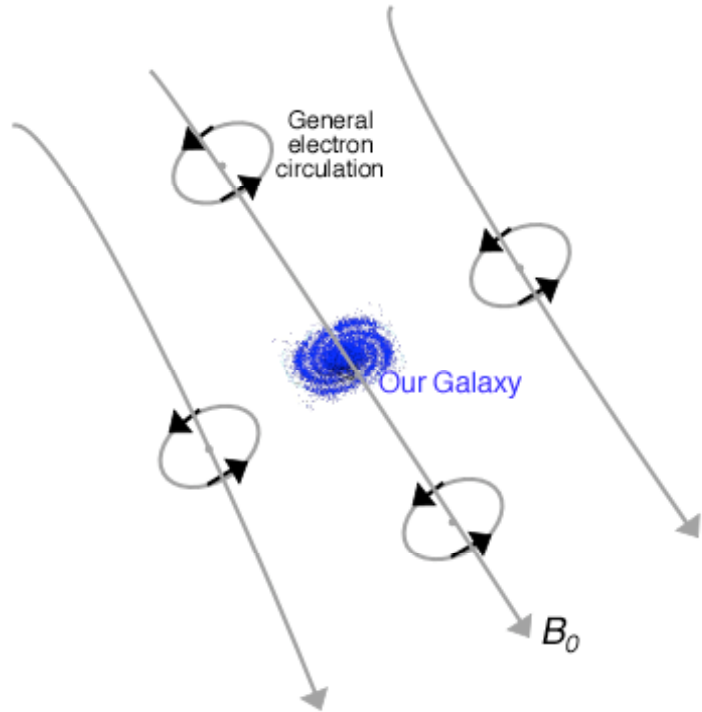

Fig. 1. If a large-scale magnetic field exists, electrons will circulate generally as shown. For electrons that are upfield and downfield from us, we see only an average transverse velocity relative to our line of sight. For electrons near 90° to the field we see a large component of velocity toward and away from us. The anisotropies in the magnetic field are mapped into the multipoles of the CMB, all of which are generally aligned in the same direction.

The mechanism proposed here is similar to the Sunyaev-Zel'dovich effect [7] which causes a change in the apparent brightness of the CMB due to inverse Compton scattering off of electrons in hot plasma inside or outside of galaxies. However, the effect proposed here can be much larger because all the electrons within the coherence length of the magnetic field contribute and its imprint covers the whole sky. The CMB will also exhibit a characteristic large-scale polarization pattern.

In a recent article, I showed that the spin orientations of spiral galaxies show an asymmetry that appears to be along the AE. [8] This analysis was based on a sample of 1660 spiral galaxies from the Sloan Digital Sky Survey (SDSS) and extended out to a redshift $z = 0.04$. The probability that this alignment occurred by chance was estimated to be <0.4%. The only plausible explanation for this alignment is that a uniform cosmic magnetic field extends over this region roughly in the direction of right ascension $(RA) \simeq 180°$ and declination $\simeq 10°$. The low-$\ell$ CMB anomalies suggest that this field extends out to the last scattering surface for the CMB around 300,000 years after the Big Bang.

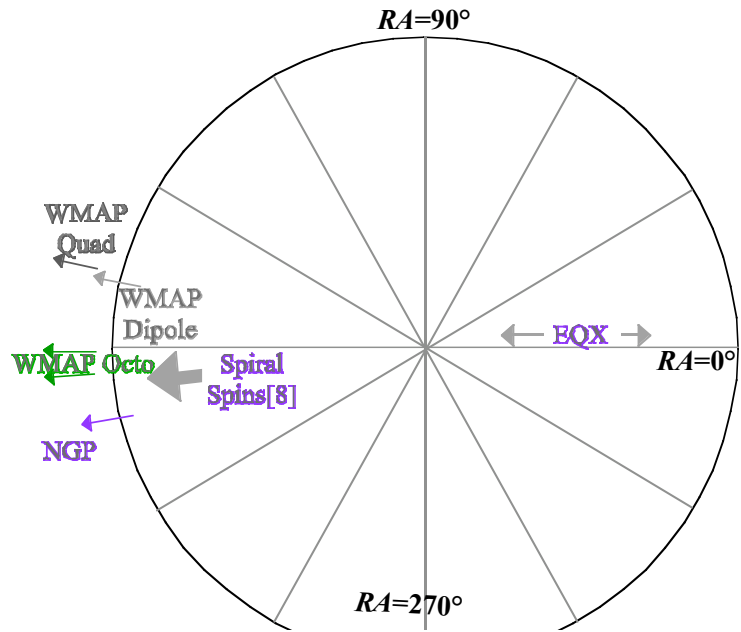

Fig. 2. Right ascensions of the alignments discussed in the text. The WMAP dipole, quadrupole, and two of the octopoles are indicated. NGP is the North Galactic pole. EQX are the equinoxes. The axis of the spiral galaxy spin asymmetry from [8] is also shown. The declinations of all these alignments are within about ±15° of each other and are ~0°.

Another surprising "coincidence" is that the poles of our Galaxy appear to be aligned with the AE. However, this is not so surprising in light of the general alignment of spiral galaxy spins along this direction [8], *i.e.,* our own Galaxy has its spin oriented along the same direction as the predominant direction for all spiral galaxies. This explains another seemingly fortuitous occurrence. – Most astronomical surveys, such as the SDSS, tend to observe generally along the direction of the Galactic poles where obscuration by the Milky Way is not a problem. Thus most of our extragalactic surveys "happen" to be close to the AE. This serendipitous bias may lead to other anomalies and must be kept in mind when interpreting extragalactic phenomena. A case in point is the supernovae data that suggest an accelerated expansion of the universe. There are no data near the plane of our Galaxy so the observations are biased along the Galactic poles. If indeed the universe is ellipsoidal with an axis oriented toward the poles, the interpretation of the data might be significantly altered.

Figure 2 shows the right ascension for all the alignments discussed here, including that of the axis of the spiral galaxy spin asymmetry from [8]. The WMAP multipoles are for the 3-year data (ILC123) as given by Copi et al. [9] All of these alignments lie close to the equinoxes. Except for one of the two WMAP octopoles, all have declinations between -7° and +16°.

In summary, I propose a simple mechanism that can account for the near alignment of the CMB low-$\ell$ multipoles. Though the dipole is generally attributed to our motion through the CMB, this mechanism could generate a dipole that is large enough to overwhelm that due to our motion. The apparently fortuitous alignment of the axis of our Galaxy with the same axis would also be explained. The worrisome alignment of the equinoxes with the AE [3] is now seen to be merely because they are defined to be at right ascensions of 180° and 0° and declination = 0°, which happen to be near the AE.

## References and Notes

1. G. Hinshaw *et al.*, astro-ph/0603451, 17 Mar 2006. Submitted to Astrophys. J.
2. M. Tegmark *et al.*, Phys. Rev. **D68**, 123523 (2003).
3. D. Schwarz *et al.*, Phys. Rev. Lett. **93**, 221301 (2004).
4. K. Land and J. Magueijo, Phys. Rev. Lett. **95**, 071301 (2005).
5. L. Campanelli, P. Cea, and L. Tedesco, Phys. Rev. Letters **97**, 131302 (2006).
6. By uniform in this context we mean uniform on very large scales far from galaxies. As galaxies form, the overall magnetic field will be greatly distorted around them.
7. R. Sunyaev and R., Ya. B. Zel'dovich, Astrophys. Space Sci. **4**, 301 (1969).
   R. Sunyaev and R., Ya. B. Zel'dovich, Comm. Astr. Space Phys. **4**, 173 (1972).
8. M. J. Longo, astro-ph/0703325v3, 13 Mar 2007. Submitted to Phys. Rev. Lett.
9. C. Copi et al., astro-ph/0605135 v2, 31 May 2006.